# Phase-stepping algorithms for synchronous demodulation of nonlinear phase-shifted fringes

MANUEL SERVIN,[*] MOISES PADILLA, IVAN CHOQUE, AND SOTERO ORDONES

*Centro de Investigaciones en Optica A. C. Loma del Bosque 115, Lomas del Campestre,37000 Leon Guanajuato, Mexico.*
*[*] mservin@cio.mx*

**Abstract:** Standard phase-stepping algorithms (PSAs) estimate the measuring phase of linear-carrier temporal-fringes with respect to a linear-reference. Linear-carrier fringes are normally obtained using feedback, closed-loop, optical phase-shifting devices. On the other hand, open-loop, phase-shifting devices, usually give fringe patterns with nonlinear phase-shifts. The Fourier spectrum of linear-carrier fringes is composed by Dirac deltas only. In contrast, nonlinear phase-shifted fringes are wideband, spread-spectrum signals. It is well known that using linear-phase reference PSA to demodulate nonlinear phase-shifted fringes, one obtains an spurious-piston. The problem with this spurious-piston, is that it may wrongly be taken as a real optical thickness. Here we mathematically find the origin of this spurious-piston and design nonlinear phase-stepping PSAs to cope with open-loop, nonlinear phase-shifted interferometric fringes. We give a general theory to tailor nonlinear phase-stepping PSAs to demodulate nonlinear phase-shifted wideband fringes.

## 1. Introduction

Linear-reference phase-shifting algorithms (PSAs) have been used to demodulate linear-carrier temporal fringes since the pioneering work by Bruning et al. [1,2]. To generate linear phase-shifting fringes one normally uses well-calibrated, feedback closed-loop, optical phase-shifters [1,2]. In contrast if one uses open-loop, phase-shifters, one normally obtain wideband, nonlinear-carrier fringes [3-9]. In these cases the PSA must also be wideband to deal with highly nonlinear phase-shifted fringes [3-9]. Hibino et al. indicated that an artifact spurious piston appears in the estimated phase when using a linear-reference PSA to demodulate nonlinear phase-shifted fringes [4-9]. This spurious-piston is a numeric artifact of the linear-reference PSA, which may be wrongly interpreted as physical optical thickness [4-9]. Real optical thickness measuring, is fundamental when testing optical material slabs in semiconductor and display equipment [4-9]. Many systematic errors have been solved in linear-carrier, linear-reference PSAs, such as phase-shift miscalibration, fringe harmonics, experiment vibrations [2-9]. For precision thickness measurement, and nonlinear phase-shifted fringes, several linear-reference PSAs with no spurious-piston have been proposed [4-9]. Recently Kim et al. have pointed-out [9] that this numerical spurious-piston has received little attention because it does not give a waving profiling error (such as detuning or harmonics) when an optical surface is profiled. However, when the central interest is to measure absolute optical thickness of transparent slabs by wavelength-tuning (for example), this numerical-piston translates into errors in thickness [4-9]. This error is given by the product of the demodulated-phase and the synthetic wavelength, which is much greater than the wavelength used [9]. Linear-reference PSAs for demodulating wideband, nonlinear-carrier fringes, have been developed using the Taylor series expansion of the arc-tangent of the phase-error [4-9]. The linear-reference PSA's coefficients are then calculated to set the first terms of this Taylor expansion to zero [4-9].

In this work we are proposing a different approach for phase demodulating temporal nonlinear-carrier fringes using wideband, synchronous, nonlinear-reference PSA. This is similar to the theory behind chirp-carrier radars [10,11]. In chirp-radars the wideband radio-frequency (RF) pulse varies quadratically with time. When the RF chirp-pulse bounce back from the radar target, the incoming RF-signal is correlated with a synchronous, local chirp-waveform. In the case of wideband chirp-radar, one is interested in timing the amplitude of the correlation peak between the incoming RF chirp-signal and the chirped local-oscillator. Timing this correlation peak give us the round-trip target distance [10,11].

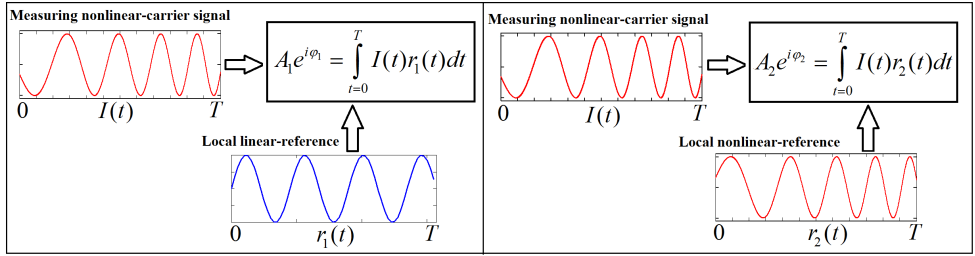

Fig.1. Schematic for linear $r_1(t)$ and nonlinear phase-shifted reference $r_2(t)$. The nonlinear phase-shifted fringes is $I(t)$. We are showing the real part of the complex-valued reference.

As Fig. 1 shows, here we are using the same concept of synchronously following the nonlinear phase-shifted carrier fringes using the same nonlinearity phase-shifting as reference.

## 2. Linear and nonlinear phase-shifted interferometric fringes

Let us first show the usual mathematical models for linear and nonlinear phase-shifted interferometric fringes. The model for linear-carrier interferometric fringes is,

$$I_1(t) = a + b\cos(\varphi + \omega_0 t); \qquad \omega_0 \in (0,\pi); \quad t \in [0,T]. \tag{1}$$

Where $\varphi \in [-\pi,\pi]$ is the measuring phase. On the other hand, nonlinear-carrier fringes are formalized by,

$$I_2(t) = a + b\cos[\varphi + \omega_0 t + \Delta(t)]; \quad \omega_0 \in (0,\pi); \quad t \in [0,T]. \tag{2}$$

We are assuming that the nonlinearity $\Delta(t)$ is smooth, and can be determined experimentally [3-9]. Previous papers assume that $\Delta(t)$ can be approximated by few Taylor series terms [3,9]. Here we relax this condition, by requiring only that the derivative of $[\omega_0 t + \Delta(t)]$ be bounded within the open interval $(0,\pi)$,

$$\left[\omega_0 + \frac{d\Delta(t)}{dt}\right] \in (0,\pi) \; ; \quad t \in [0,T] \; . \tag{3}$$

In Fig. 2 we show linear (in blue) and nonlinear (in red) phase-shifted fringes.

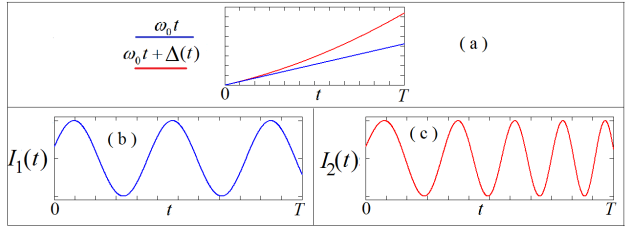

Fig. 2. Panel (a) shows in blue, linear phase-shifting, and in red, nonlinear phase-shifting. Panel (b) shows linear-carrier fringes. Panel (c) shows nonlinear-carrier fringes.

As we prove in the next sections, the nonlinear phase-shifting $\Delta(t)$ generate a spurious-numerical piston when a linear-reference PSA is used as phase-demodulator [4-9]. Fig.2 shows an example of linear and nonlinear carrier fringes,

### 3. Fourier spectrum for linear and nonlinear phase-shifted fringes

From Eq. (1), linear fringes are single-frequency at $\omega_0$, having a spectrum given by [2],

$$F\left\{a + \frac{b}{2}e^{i[\varphi+\omega_0 t]} + \frac{b}{2}e^{-i[\varphi+\omega_0 t]}\right\} = a\delta(\omega) + \frac{b}{2}e^{i\varphi}\delta(\omega-\omega_0) + \frac{b}{2}e^{-i\varphi}\delta(\omega+\omega_0). \qquad (4)$$

Where $F[\cdot]$ is the Fourier transform operator (see Fig. 3(a)). In contrast, highly nonlinear phase-shifted fringes (Eq. (2)) are wideband, and its spectrum may be modeled as,

$$F\left\{a + \frac{b}{2}e^{i\varphi}e^{i[\omega_0 t+\Delta(t)]} + \frac{b}{2}e^{-i\varphi}e^{-i[\omega_0 t+\Delta(t)]}\right\} = a\delta(\omega) + \frac{b}{2}e^{i\varphi}C(\omega) + \frac{b}{2}e^{-i\varphi}C^*(-\omega), \qquad (5)$$

Where,

$$C^*(-\omega) = F\left\{e^{-i[\omega_0 t+\Delta(t)]}\right\}; \quad C(\omega) = F\left\{e^{i[\omega_0 t+\Delta(t)]}\right\}. \qquad (6)$$

Figure 3 shows schematically the spectrum of linear and wideband nonlinear-carrier fringes. In Eq. (5) the terms $C^*(-\omega)$ and $C(\omega)$ are wideband spectra.

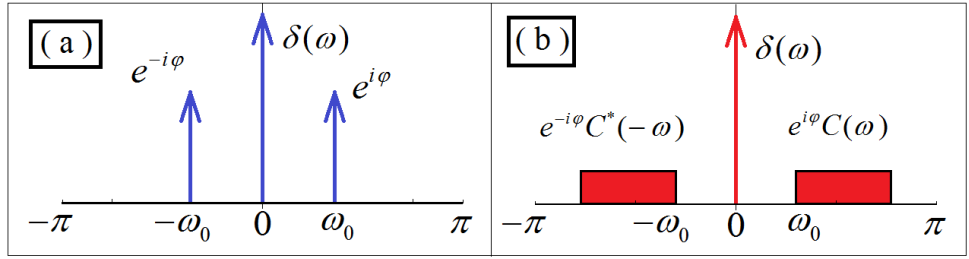

Fig. 3. Panel (a) shows the three delta-spectrum of linear-carrier fringes, and in panel (b) the wideband spectrum of nonlinear phase-shifted fringes.

Summarizing, linear-phase carrier fringes have a three delta spectrum (Fig 3(a)); while nonlinear-phase carrier fringes have two spread-spectrum components (Fig. 3(b)).

### 4. Linear and nonlinear reference PSAs

Let us now show the mathematical form of phase-shifting algorithms (PSAs) using linear and nonlinear-reference for demodulating nonlinear-carrier fringes

*3.1 Standard linear-reference PSA for demodulating linear-carrier fringes*

The general form for standard linear-carrier, linear-reference PSA is [2],

$$Ae^{i\varphi} = \sum_{n=0}^{N-1} \overbrace{[c_n e^{i\omega_0 n}]}^{Linear\ Reference} \overbrace{[a+b\cos(\varphi+\omega_0 n)]}^{Linear\ Carrier\ Fringes}. \qquad (7)$$

These are the standard linear-carrier, linear-reference PSAs in use since 1974 [1,2].

*3.2 Linear-reference PSA for demodulating nonlinear-carrier fringes*

People has proposed linear-reference PSAs to demodulate nonlinear-carrier fringes as [3-9],

$$A_1 e^{i(\varphi + Piston)} = \sum_{n=0}^{N-1} \overbrace{\left[ d_n e^{i\,\omega_0 n} \right]}^{\text{Linear Reference}} \overbrace{\left[ a + b\cos\left( \varphi + \omega_0 n + \Delta(n) \right) \right]}^{\text{Nonlinear Carrier Fringes}}. \tag{8}$$

As we show next, using a linear-reference PSA, we generally obtain a spurious-piston, $Piston \neq 0$ [4-9]. Hibino et al., have proposed linear-reference PSAs to eliminate this spurious piston [4-9]. Here we are proposing an alternative solution, a more natural way (we believe), for disappearing this non-desired, spurious-piston.

*3.3 Nonlinear-reference PSA for demodulating nonlinear-carrier fringes*

We specifically propose the use of a nonlinear-reference PSA which has the following form,

$$A_2 e^{i\varphi} = \sum_{n=0}^{N-1} \overbrace{\left[ w_n e^{i[\omega_0 n + \Delta(n)]} \right]}^{\text{Nonlinear Reference}} \overbrace{\left[ a + b\cos\left( \varphi + \omega_0 n + \Delta(n) \right) \right]}^{\text{Nonlinear Carrier Fringes}}; \quad (w_n \in \mathbb{R}). \tag{9}$$

Note that the nonlinear-reference $\exp[i(\omega_0 n + \Delta(n))]$ is synchronous with the nonlinear-carrier $\cos[\varphi + \omega_0 n + \Delta(n)]$; this fact makes the spurious-piston disappear $Piston=0$. The weighting coefficients $(w_n)$ are chosen to approximate a Hilbert quadrature filter.

## 5. Spurious-piston using linear phase-shifted reference PSAs

Using a linear-reference PSAs to demodulate nonlinear-carrier fringes (Eq. (8)) one obtains,

$$A_1 e^{i[\varphi + Piston]} = \sum_{n=0}^{N-1} d_n e^{-i\,\omega_0 n} I_2(n) = \sum_{n=0}^{N-1} \left\{ a + \frac{b}{2} e^{i\varphi} e^{i[\omega_0 n + \Delta(n)]} + \frac{b}{2} e^{-i\varphi} e^{-i[\omega_0 n + \Delta(n)]} \right\} d_n e^{-i\,\omega_0 n} \tag{10}$$

Performing the indicated multiplications one obtains,

$$A_1 e^{i[\varphi + Piston]} = a\left[\sum_{n=0}^{N-1} d_n e^{-i\,\omega_0 n}\right] + \frac{b}{2} e^{i\varphi} \left\{\sum_{n=0}^{N-1} d_n e^{i\Delta(n)}\right\} + \frac{b}{2} e^{-i\varphi} \left[\sum_{n=0}^{N-1} d_n e^{-i[2\omega_0 n + \Delta(n)]}\right] \tag{11}$$

The coefficients $d_n$ are chosen such that the first and third square-brackets are set to zero as,

$$\left[\sum_{n=0}^{N-1} d_n e^{-i\,\omega_0 n}\right] = 0, \quad \text{and} \quad \left[\sum_{n=0}^{N-1} d_n e^{-i[2\omega_0 n + \Delta(n)]}\right] = 0. \tag{12}$$

Obtaining the desired analytic signal as,

$$A_1 e^{i(\varphi + Piston)} = \frac{b}{2} e^{i\varphi} \left\{\sum_{n=0}^{N-1} d_n e^{i\Delta(n)}\right\}; \quad Piston = \arg\left\{\sum_{n=0}^{N-1} d_n e^{i\Delta(n)}\right\}. \tag{13}$$

As we see, in general, the spurious-piston is non-zero ( $Piston \neq 0$ ), and it may give erroneous absolute optical thickness measurements [4-9].

## 6. No spurious-piston using nonlinear phase-shifted reference PSA

Now using a synchronous (matched-phase) nonlinear reference PSA (Eq. (9)) one gets,

$$A_2 e^{i\varphi} = \sum_{n=0}^{N-1} w_n e^{-i[\omega_0 n + \Delta(n)]} I_2(n) = \sum_{n=0}^{N-1} \left[ a + \frac{b}{2} e^{i\varphi} e^{i[\omega_0 n + \Delta(n)]} + \frac{b}{2} e^{-i\varphi} e^{-i[\omega_0 n + \Delta(n)]} \right] w_n e^{-i[\omega_0 n + \Delta(n)]} \tag{14}$$

Performing the multiplications one obtains,

$$A_2 e^{i\varphi} = a\left[\sum_{n=0}^{N-1} w_n e^{-i[\omega_0 n + \Delta(n)]}\right] + \frac{b}{2} e^{i\varphi} \left[\sum_{n=0}^{N-1} w_n\right] + \frac{b}{2} e^{-i\varphi} \left[\sum_{n=0}^{N-1} w_n e^{-i[2\omega_0 n + \Delta(n)]}\right] \tag{15}$$

As we did before, to obtain just the desired analytic-signal $\exp[i\varphi]$ one needs,

$$\left[\sum_{n=0}^{N-1} w_n e^{-i[\omega_0 n + \Delta(n)]}\right] = 0; \quad and \quad \left[\sum_{n=0}^{N-1} w_n e^{-i[2\omega_0 n + \Delta(n)]}\right] = 0. \quad (16)$$

Obtaining the phase-demodulated signal $Ae^{i\varphi}$ as,

$$A_2 e^{i\varphi} = \frac{b}{2} e^{i\varphi} \sum_{n=0}^{N-1} w_n \; ; \quad (w_n \in \mathbb{R}). \quad (17)$$

The spurious piston has naturally disappeared (*Piston*=0) thanks to the use of a synchronous nonlinear-reference $\exp\{-i[\omega_0 n + \Delta(n)]\}$ in the PSA.

## 7. Spectral design for nonlinearly phase-shifted reference PSAs

In previous section we gave an algebraic approach for calculating the coefficients $(w_n \in \mathbb{R})$ for nonlinear phase-shifted reference PSAs. Here we develop a more intuitive spectral design. The impulse response of the nonlinear reference PSA (Eq. (9)) is.

$$h_2(t) = \sum_{n=0}^{N-1} w_n e^{i[\omega_0 n + \Delta(n)]} \delta(t-n); \quad (w_n \in \mathbb{R}). \quad (18)$$

Then its FTF is $H_2(\omega) = F[h_2(t)]$,

$$H_2(\omega) = F\left\{\sum_{n=0}^{N-1} w_n e^{i[\omega_0 n + \Delta(n)]} \delta(t-n)\right\} = \sum_{n=0}^{N-1} w_n e^{i[\omega_0 n + \Delta(n)]} e^{-in\omega}. \quad (19)$$

The coefficients $(w_n)$ of $H_2(\omega)$ are chosen to fulfill with the wideband conditions,

$$\begin{aligned} H_2(\omega) &= 0 \quad for \quad \omega \in [-\pi, 0] \\ H_2(\omega) &\neq 0 \quad for \quad \omega \in (0, \pi) \end{aligned}. \quad (20)$$

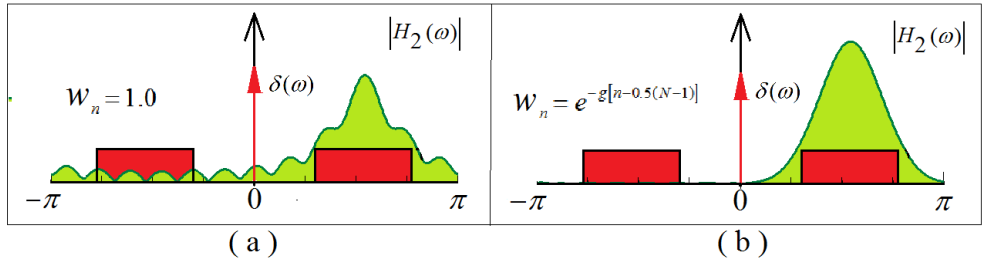

Fig. 4. Two nonlinear-reference PSA, FTF-spectra. Panel (a) shows (in green) the FTF of a square-window, nonlinear-reference PSA. Panel (b) shows the FTF of a nonlinear-reference PSA with Gaussian window [12,13]. In red we show the wideband spectrum of the nonlinear-carrier fringes. The FTF in panel (b) is a smooth approximation of a Hilbert quadrature filter.

As Fig. 4(a) shows $H_2(\omega)$ for a square-window $(w_n = 1.0)$. We can see that $H_2(\omega)$ is not zero for $\omega \in [-\pi, 0]$, obtaining an erroneous phase. One solution to this is to use an apodizing window [12,13]. We used a Gaussian window $w_n = e^{-g[n-0.5(N-1)]^2}; (g < 1.0)$, and its FTF $H_2(\omega) = F[h_2(t)]$ is shown in Fig. 4(b). Of course other weightings windows $(w_n \neq 1)$ may be used [12,13].

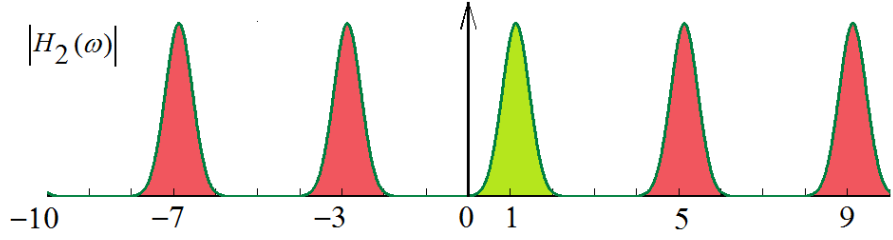

Fig. 5. Fundamental (in green) and harmonic (in red) FTF response for the nonlinear-reference PSA. As usual the harmonics {-7,-3,5,9} response is asymmetrical.

In Fig. 5 we show the (normalized frequency) harmonic response of the apodized, nonlinear reference PSA.

## 8. Signal-to-noise ratio (SNR) for linear and nonlinear reference PSA

Here we find the SNR [2] of the phase-demodulated nonlinear-carrier fringes corrupted by additive white Gaussian-noise (AWGN) . The noisy fringes are,

$$I_2(t) = \sum_{n=0}^{N-1}\{ a + b\cos[\varphi + \omega_0 t + \Delta(t)] + N(t) \}\delta(t-n). \quad (21)$$

Where the noise spectral density $S(\omega)$ is flat, and it is given by,

$$S(\omega) = F[R_{NN}(\tau)] = F[E\{N(t)N(t+\tau)\}] = \frac{N_0}{2}; \quad \omega \in [-\pi,\pi]. \quad (22)$$

Being $R_{NN}(\tau) = E\{N(t)N(t+\tau)\}$ the ensemble autocorrelation function of $N(t)$ [2]. The flat noise power-spectrum of $N(t)$ is modified to $(N_0/2)|H_2(\omega)|^2$ [2].

For nonlinear-carrier fringes, and linear-reference PSA, the SNR is given by,

$$\text{SNR}_1 = \frac{\text{Signal Energy}}{\text{Noise Energy}} = \frac{\left(\dfrac{b}{2}\right)^2 \left|\sum_{n=0}^{N-1} d_n e^{-i\Delta(n)}\right|^2}{\left(\dfrac{N_0}{2}\right)\int_{-\pi}^{\pi}|H_1(\omega)|^2 d\omega}. \quad (23)$$

On the other hand for nonlinear-carrier and reference, the SNR is given,

$$\text{SNR}_2 = \frac{\text{Signal Energy}}{\text{Noise Energy}} = \frac{\left(\dfrac{b}{2}\right)^2 \left|\sum_{n=0}^{N-1} w_n\right|^2}{\left(\dfrac{N_0}{2}\right)\int_{-\pi}^{\pi}|H_2(\omega)|^2 d\omega}. \quad (24)$$

Where $(N_0/2)\int|H_1(\omega)|^2 d\omega$, and $(N_0/2)\int|H_2(\omega)|^2 d\omega$ are the total filtered-noise energy. The energy of $A_2 e^{i\varphi}$ using the nonlinear-reference PSA is generally higher than the energy of $A_1 e^{i[\varphi+\text{Piston}]}$ using a linear-reference PSA, this is because,

$$\left(\frac{b}{2}\right)^2 \left|\sum_{n=0}^{N-1} w_n\right|^2 \geq \left(\frac{b}{2}\right)^2 \left|\sum_{n=0}^{N-1} d_n e^{-i\Delta(n)}\right|^2. \quad (25)$$

Assuming that both have about the same bandwidth $|H_1(\omega)| \approx |H_2(\omega)|$, one obtains,

$$\text{SNR}_2 \geq \text{SNR}_1. \tag{26}$$

As conclusion, the SNR is generally higher for a PSA with nonlinear-reference.

## 9. Example of a 13-steps Gaussian-window nonlinear-reference PSA

Here we are given a computer simulation example of a 13-step nonlinear-reference PSA applied to nonlinear-carrier fringes. The most usual phase-shifted nonlinearity is quadratic, $\Delta(n) = \varepsilon_2 n^2$ [3-9]. We start by considering nonlinear-carrier fringes as,

$$I(t;\varphi) = \sum_{n=0}^{12}\left[1 + \cos(\varphi + \omega_0 n + \varepsilon_2 n^2)\right]\delta(t-n); \quad (\omega_0 = 0.35\pi, \varepsilon_2 = 0.05\omega_0). \tag{27}$$

The non-linear phase and the interferometric chirp-waveform is shown in Fig. 6.

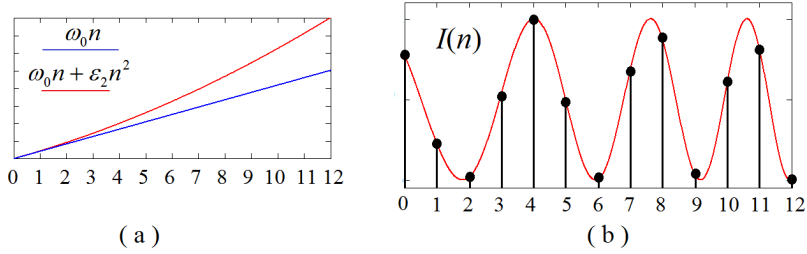

Fig. 6. Panel (a) shows the linear (in blue) and a strong nonlinear, quadratic-carrier (in red). Panel (b) shows the chirp-carrier fringes sampled at a constant sampling rate.

The specific form of the 13-steps nonlinear, chirp-reference PSA is given by,

$$A_2 e^{i\varphi} = \sum_{n=0}^{12} w_n e^{i\left[\omega_0 n + \varepsilon_2 n^2\right]} I(n;\varphi); \quad w_n = e^{-0.1\left[n - 0.5*(N-1)\right]^2}. \tag{28}$$

We have assumed no linear detuning. The PSA reference $\exp\{-i[\omega_0 n + \varepsilon_2 n^2]\}$, is synchronous with the chirp-fringes. The FTF ($H_2(\omega)$) of the nonlinear-reference PSA is,

$$H_2(\omega) = F[h_2(t)] = F\left[\sum_{n=0}^{12} w_n e^{i\left[\omega_0 n + \varepsilon_2 n^2\right]}\delta(t-n)\right] = \sum_{n=0}^{12} w_n e^{i\left[\omega_0 n + \varepsilon_2 n^2\right]} e^{-in\omega}. \tag{29}$$

And its temporal and spectral graphs are shown in Fig. 7.

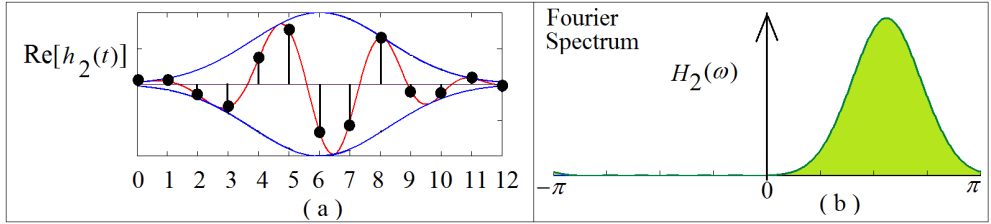

Fig. 7. Panel (a) shows the real component of the complex-valued chirp-wavelet impulse response. In panel (b) we show the PSA-FTF of the chirp-wavelet impulse response. The weighting coefficients shape the FTF as a smooth approximation of a Hilbert quadrature filter.

Next Fig. 8 shows, superimposed, the fringe-data and the chirp-reference PSA spectra.

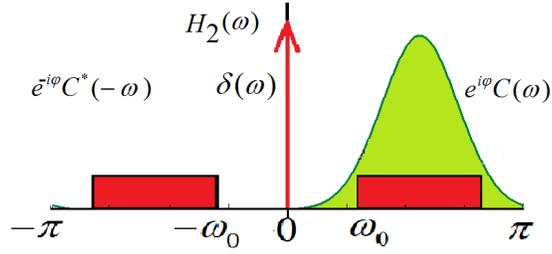

Fig. 8. Superimposed nonlinear fringe spectrum (in red), and FTF of the Gaussian-window, nonlinear-reference PSA (in green). This FTF has negligible response at the negative frequencies of the wideband fringes. This FTF approximates a Hilbert quadrature filter.

We evaluate our nonlinear-reference PSA demodulation error by the following formula,

$$\varphi_{Error} = \varphi - \arg\left[\sum_{n=0}^{12} w_n e^{i(\omega_0 n + \varepsilon_2 n^2)} I(n;\varphi)\right]; \quad \varphi \in [0, 2\pi]. \tag{30}$$

Finally Fig. 9 shows the phase estimation error $\varphi_{Error}$, for $\varphi \in [0, 2\pi]$.

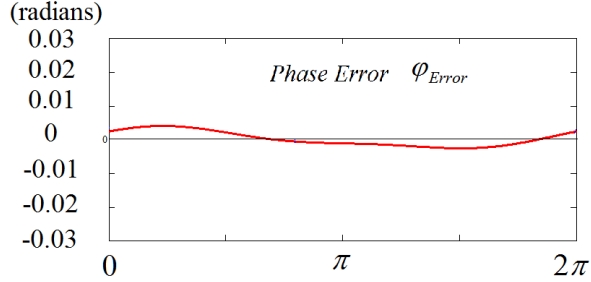

Fig. 9 Phase error given by Eq. (32). Note the vertical scale is within [-0.03,0.03] radians.

Figure 9 shows that the peak phase demodulation error $\varphi_{Error}$ is about 0.04 radians.

**10. Example of a 13-steps square-window nonlinear-reference PSA**

Here we analyze a square-window nonlinear-reference PSA for the same fringes used in previous section $\Delta(n) = \varepsilon_2 n^2$. Then our square-window PSA is,

$$A_2 e^{i\varphi_{Square}} = \sum_{n=0}^{12} w_n e^{i\left[\omega_0 n + \varepsilon_2 n^2\right]} \left\{1 + \cos\left[\varphi + \omega_0 n + \varepsilon_2 n^2\right]\right\}; \quad w_n = 1.0. \tag{31}$$

The spectral graph of the FTF associated to this PSA is shown in Fig. 10.

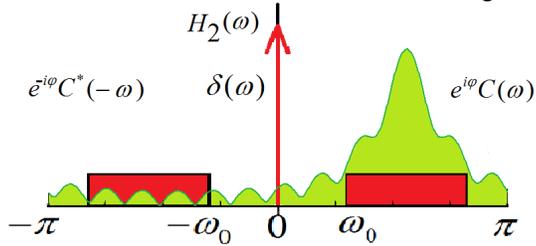

Fig. 10. Spectral response (FTF) for the square-window, nonlinear-reference PSA. This square window cannot be used because it has large response in the origin and the left side of the fringe spectrum. This FTF is a bad approximation of a one-sided Hilbert quadrature filter.

As Fig. 11 shows, the DC background of the fringes is not fully filtered-out, and also large energy from the unwanted conjugate-signal leaks into the desired analytic signal.

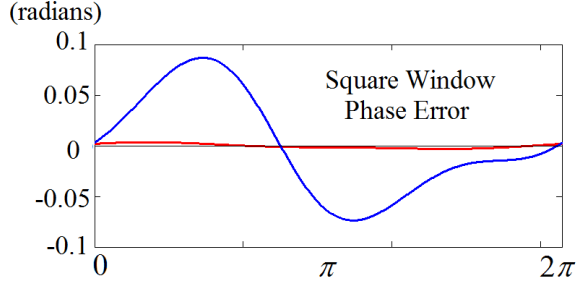

Fig. 11. The blue trace shows the phase-error for the 13-step, square-window, nonlinear-reference PSA. For comparison, the red-trace is the phase error corresponding to the Gaussian window seen in previous section. Note that the vertical scale is now [-0.1,0.1] radians.

Figure 11 shows in the blue trace the phase-error of the square-window nonlinear-reference PSA. We summarize this section by remarking the fact that synchronously following the nonlinear-carrier variations of the fringes is not enough. One must also apply an apodizing weighting window to the nonlinear-reference PSA [12,13].

## 11. Discussion of the proposed nonlinear-reference PSA theory

Before our general summary, we want to make a clear wrap-up of our contribution. The basic theory presented herein is completely general. The only restriction is that the nonlinear-carrier fringe spectrum be bandlimited. For the readers' convenience we rewrite the few equations required. The nonlinear-carrier fringes are modeled as,

$$I_2(t) = a + b\cos[\varphi + \omega_0 t + \Delta(t)]; \quad t \in [0,T]; \quad \omega_0 \in (0,\pi). \tag{32}$$

Where the *only* restriction about the phase-shifted variation $[\omega_0 t + \Delta(t)]$ is,

$$\left[\omega_0 + \frac{d\Delta(t)}{dt}\right] \in (0,\pi) \; ; \quad t \in [0,T] . \tag{33}$$

This restriction on $\Delta(t)$ is more general to previous efforts which assumed that $\Delta(t)$ should be expansible as a Taylor series [3-9]. We then proposed our nonlinear-reference PSA as,

$$A_2 e^{i\varphi} = \sum_{n=0}^{N-1} w_n e^{i[\omega_0 n + \Delta(n)]} I_2(n); \quad (w_n \in \mathbb{R}). \tag{34}$$

The window ($w_n$) shape the FTF of this PSA, and it may be taken from a wide set of functions [12,13]. This analytic signal has no spurious piston. The FTF of this PSA is,

$$H_2(\omega) = F[h_2(t)] = F\left\{\sum_{n=0}^{N-1} w_n e^{i[\omega_0 n + \Delta(n)]} \delta(t-n)\right\} = \sum_{n=0}^{N-1} w_n e^{i[\omega_0 n + \Delta(n)]} e^{-in\omega} . \tag{35}$$

The coefficients ($w_n$) shape this FTF to cover the right hand-side spectrum of the fringes as,

$$\begin{array}{ll} H_2(\omega) = 0 & for \quad \omega \in [-\pi, 0] \\ H_2(\omega) \neq 0 & for \quad \omega \in (0, \pi) \end{array} . \tag{36}$$

These five equations succinctly show our theory for designing nonlinear-reference PSAs for phase-demodulating nonlinear-carrier fringes without spurious-piston.

## 12. Summary

Here we have given a frequency transfer function (FTF) approach for designing nonlinear phase-shifting algorithms (PSAs) applied to demodulate nonlinear phase-shifted fringes. The estimated nonlinear phase-step variations of the fringes [4-9], constitute also our nonlinear phase-step reference for the PSA. We then find the real-valued PSA coefficients ($w_n$) that shapes the FTF spectrum of the PSA. The spectral shape of the nonlinear reference PSA smoothly approximate a Hilbert quadrature filter. As such, the spectral FTF shaping must render almost zero the left side (including zero) of the fringes spectrum.

Phase-demodulation of nonlinear phase-shifted fringes using linear-reference PSAs has been studied before [3-9]. As mentioned, in general, linear-reference PSAs obtain an artifact, numeric piston when demodulating nonlinear-carrier fringes [4-9]. This spurious-piston may render the phase measurement of absolute optical thickness erroneous [4-9]. Hibino et al have eliminated this spurious-piston by imposing some conditions on the coefficients of linear-reference PSAs [4-9]. Here we have seen that using a nonlinear-reference PSA synchronous with the nonlinear-carrier fringes, the spurious piston disappears in a more natural way (we believe). We think that our nonlinear-phase reference PSA approach shed new light for understanding temporal phase-demodulation of broadband interferometric fringes.